\documentclass{article}

\title{Covariant approach to equilibration in effective field theories}
\author{Mark Burgess\\Faculty of Engineering\\Oslo University College\\0254 Oslo,
Norway\\~\\M.E. Carrington \\Department of Physics\\Brandon University\\Manitoba, Canada R7A 6A9\\~\\Gabor Kunstatter\\Department of
Physics\\University of Winnipeg\\Manitoba, Canada R3B 2E9}
\date{\today}

\def\beq{\begin{eqnarray}}
\newcommand{\xbar}{\bar x}
\def\eeq{\end{eqnarray}}
\def\2{\frac{1}{2}}
\def\boxx{{\vcenter{\vbox{\hrule height.3pt
          \hbox{\vrule width.3pt height6pt
          \kern6pt\vrule width.3pt}\hrule height.3pt}}\;}}

\usepackage{a4}
\begin{document}
\maketitle

\begin{abstract}
The equilibration of two coupled reservoirs is studied using a Green
function approach which is suitable for future development with the
closed time path method. The problem is solved in two
parameterizations, in order to demonstrate the non-trivial issues of
parameterization in both the intermediate steps and the interpretation
of physical quantities. We use a covariant approach to find
self-consistent solutions for the statistical distributions as
functions of time.  We show that by formally introducing covariant
connections, one can rescale a slowly varying non-equilibrium theory
so that it appears to be an equilibrium one, for the purposes of
calculation.  We emphasize the importance of properly tracking
variable redefinitions in order to correctly interpret physical
quantities.
\end{abstract}

\section{Introduction}

Effective field theory is an increasingly important tool in the
repertoire of theoretical physics. The problem of the equilibration of
statistical systems has been studied in many contexts
\cite{reif1,schwinger2,feynman,cal,calzetta1,calzetta2}, 
but has not been formulated in the
language of effective quantum field theories, where
re-parameterizations of basic quantities can obscure the physics of
the process.

Suppose we consider two systems with different mechanical and
thermodynamical properties and couple these together. What is the
outcome?  Intuition tells us that such a system must achieve a state
of thermodynamical equilibrium. However, it is not clear how this
equilibrium is achieved using the language of effective field theory.
In terms of particles, it is straightforward to picture the exchange
of energy between heat baths: there is a physical flow of point-like
objects carrying momentum, which are exchanged until, eventually,
there is an equal spectrum of energies on both sides of the
interface. In a field theory however, one deals with quanta or modes
of the field, so there is no direct notion of a flow of hot into cold
or vice versa, only the occupation of states in a Fock space.

The procedure of identifying a functional effective field theory
involves many variable redefinitions and partial summations. Each
re-parameterization distances the system from any intuition one might
have about its behaviour.  Our goal in this paper is to understand the
process of equilibration in the language of field theory in the
context of a specific model. We consider a pair of essentially free
scalar fields with different masses and different initial temperatures
which are coupled minimally at an initial time. For definiteness we
assume that the systems are in direct contact, rather than interacting
through a boundary layer, so that the two systems essentially occupy
the same space. Although it is convenient to focus on a specific
model, our aim is to extract general principles about non-equilibrium
field theory.  We employ the methods of general covariance during
field redefinitions, in order to track the meaning of the quantities
which arise in the formulation, and also to find the optimal
parameterization for solving the equations. As is often the case with
covariance, this leads to important physical insights.

The notation and conventions are those of reference
\cite{burgess12,burgess13}, with $n$ spatial dimensions and one time
dimension. To simplify notation we shall use Schwinger's shorthand for
the measures:
\beq
\int \frac{d^{n+1}k}{(2\pi)^{n+1}} &\equiv& (dk)\nonumber\\
\int \frac{d^{n}{\bf k}}{(2\pi)^{n}} &\equiv& (d{\bf k})\nonumber\\
dV_x=dtd^n{\bf x} &\equiv& (dx)
\eeq
and so on.

A non-equilibrium system remembers its history; it is not translationally
invariant in time. It is convenient to parameterize Green functions in terms
of even and odd variables over any interval from $x$ to $x'$:
\beq
\tilde x &=& (x-x')\nonumber\\
\overline x &=& \2 (x+x').
\eeq
The odd variables $\tilde x$ characterize translational degrees of
freedom, while the even variables $\overline x$ represent
inhomogeneous impediments to translational freedom, or alternatively,
characterize local curvature in the dynamical and kinematical
parameters.

The formulation we adopt is based on \cite{burgess12}, which is built
around Schwinger's closed time path quantum action principle. We do
not need to formulate the full closed time path problem here, but our
approach is easily generalizable for future problems. We concentrate
on understanding the basic unitary processes which act in the system,
and employ the language of sources and Green functions to describe the
development.

\section{The model}

Consider two systems which are brought into contact, so as to interact
through a minimal coupling which is not manifestly $Z_2$
($\phi\rightarrow-\phi$) invariant.  Once coupled, the systems
influence one another. Since the systems are statistical mixtures of
field modes, this influence can be viewed at two levels: as the
microscopic interaction of individual field solutions between the two
systems, or as a change in the statistical averages of the two systems
as a result of the microscopic interaction. It is at this latter level
that one expects to achieve equilibrium.  The model we choose begins
as a free field theory, composed of a thermal mixture of quanta. This
initial state is clearly idealized since one normally requires some
form of interaction in order to achieve thermal equilibrium in the
first place. In the interests of simplicity, we suppress all such
non-linear interactions and assume that the initial state is two
effectively free thermal baths.

We formulate the physical system using the Schwinger action
principle. The uncoupled systems have the action operators:
\beq
S_1 &=& \int (dx) \left\{(\partial_\mu\hat\Phi_1)^*(\partial^\mu\hat\Phi_1)+\hat m_1^2 \right\}\nonumber\\
S_2 &=& \int (dx) \left\{(\partial_\mu\hat \Phi_2)^*(\partial^\mu\hat \Phi_2)+\hat m_2^2\right\}.
\eeq
Before coupling the equations of motion give us the eigenvalue equation:
\beq
\hat{\cal O} \hat \Phi=0 ~~~ {\rm or} ~~~ 
\left( 
\begin{array}{cc}
-\boxx+\hat m_1^2 & 0\\
0 & -\boxx +\hat m_2^2
\end{array}
\right)
\left( 
\begin{array}{c}
\hat\Phi_1\\
\hat\Phi_2
\end{array}
\right)
 = 0.
\eeq

The Green function $\hat {\cal G}$ for the operator $\hat {\cal O}$
satisfies the equation $\hat{\cal O} \hat{\cal G} = I$. We write the
Green function as the sum of a particular integral and a complementary
function
\beq
\hat{\cal G} =  \hat G + 2\pi i\hat {\cal F}
\eeq
which satisfy
\beq
\hat {\cal O} \hat G = I\,;~~~~\hat {\cal O} \hat {\cal F} = 0\,.
\eeq
In most pure state cases one sets $\hat {\cal F}=0$ and chooses a
special solution for the Green function $G$ which satisfies certain
specific boundary conditions (retarded, advanced, etc.).  The
complementary function $\hat {\cal F}$ becomes important in
statistical many particle systems where mixtures of fields are
significant. However, it is not possible to associate specific
boundary conditions with the free particle solutions that are
contained in the complementary piece (free particle solutions do not
respect time-specific boundary conditions, since plane-wave solutions
have no beginning or end). The coefficients of plane wave solutions
must therefore change in time in order to follow the development of
the system, in a non-equilibrium many particle system. This time
dependence is discussed in detail in the next sections.  We obtain,
\beq
\hat G = \left( 
\begin{array}{cc}
\frac{1}{-\boxx+\hat m_1^2}& 0\\
0 & \frac{1}{-\boxx+\hat m_2^2}
\end{array}
\right)
\eeq
and
\beq
\hat {\cal F} = \left( 
\begin{array}{cc}
\hat f_{1}(k_0)\delta(\hat\chi_1) & 0\\
0 &\hat f_{2}(k_0)\delta(\hat\chi_2)
\end{array}
\right).\label{d1}
\eeq
Although nothing formally precludes off diagonal terms in $\hat {\cal
F}$ where the upper right contribution is proportional to
$\delta(\hat\chi_1)$ and the lower left contribution is proportional
to $\delta(\hat\chi_2)$, the boundary conditions tell us that there is
no interaction between the two systems, so off-diagonal terms would
not make sense.

Once coupled, the fields are influenced by their mutual
interaction. Since this interaction leads to modifications to the
constraints (mass shell) of the systems, we denote the coupled
variables without carets:
\beq
S_1 &=& \int (dx) \left\{(\partial_\mu\Phi_1)^*(\partial^\mu\Phi_1)+\hat m_1^2 \Phi_1^*\Phi_1
+ J_1^*\Phi_1 + J_1\Phi^*_1 \right\}\nonumber\\
S_2 &=& \int (dx) \left\{(\partial_\mu\Phi_2)^*(\partial^\mu\Phi_2)+\hat m_2^2 \Phi_2^*\Phi_2
+ J_2^*\Phi_2 + J_2\Phi^*_2 \right\}
\eeq
Both systems interact only through linear perturbations.  In order to
couple the systems together we choose:
\beq
J_1 &=&  m_p^2\;\Phi_2\nonumber\\
J_2 &=&  m_p^2\;\Phi_1,
\eeq
where $m_p^2$ is a squared polaron mass introduced for dimensional
consistency so that each field is a source for the other. Starting
from the total action,
\beq
S_{\rm tot} = (S_1 + S_2) \big|_{J_1= m_p^2 \Phi_2,J_2= m_p^2\Phi_1},\label{action}
\eeq
the operator equations of motion are,
\beq
(-\boxx + \hat m_1^2) \Phi_1 &=& -  m_p^2 \Phi_2\nonumber\\
(-\boxx + \hat m_2^2) \Phi_2 &=& -  m_p^2 \Phi_1.\label{eom}
\eeq
Note that the coupling destroys the diagonal structure of the
equations of motion. We have:
\beq
\label{new}
\hat{\cal O} \Phi = J \Phi  
\eeq
With ${\cal O} = \hat{\cal O} - J$ we can write, 
\beq
{\cal O} \Phi = 0 ~~~ {\rm or} ~~~
\left(
\begin{array}{cc}
-\boxx+\hat m_1^2 & -m_p^2\\
-m_p^2       & -\boxx+\hat m^2_2
\end{array}
\right)
\left(
\begin{array}{c}
\Phi_1\\\Phi_2
\end{array}
\right)=0.\label{eom1}
\eeq

In order to find the corresponding Green function we diagonalize the
matrix ${\cal O}$.  The technique is standard.  We solve the
eigenvalue equation
\beq
\left(
\begin{array}{cc}
-\boxx+\hat m_1^2 & -m_p^2\\
-m_p^2       & -\boxx+\hat m^2_2
\end{array}
\right)
\left(
\begin{array}{c}
\varphi_1\\\varphi_2
\end{array}
\right)
=
(-\boxx+m_\pm^2)
\left(
\begin{array}{c}
\varphi_1\\\varphi_2
\end{array}
\right).\label{det}
\eeq
and compute the invertible unitary matrix of eigenvectors $U$. The diagonalized matrix is obtained from the similarity transformation,
\beq
{\cal O}' = (U^\dagger)^{-1} {\cal O} U^{-1} = U {\cal O} U^{-1} \label{Otrans}
\eeq
The Green function matrix transforms according to the same rule as the
operator itself:
\beq
{\cal G}\rightarrow {\cal G}' = U\,{\cal G}\,U^{-1}.
\eeq
which preserves the relationship
\beq
{\cal O} {\cal G} ={\cal O}' {\cal G}' = I
\eeq
The similarity transformation corresponds to a field transformation of the form,
\beq
\Phi_A = U_{AB}\;\varphi_B, \label{rot}
\eeq
where $A,B=1,2$. 

We proceed as follows.  Solving the eigenvalue equation (\ref{det}) for the eigenvalues gives,
\beq
m^2_\pm &=& \overline m^2 \pm \sqrt{\tilde m^4 + m_p^4}\nonumber\\
&\equiv& \overline m^2 \pm m_R^2\,.\label{neweigen}
\eeq
with
\beq
\bar m^2 := \frac{1}{2}(\hat m_1^2 + \hat m_2^2) \,;~~~~\tilde m^2 := \frac{1}{2}(\hat m_1^2 - \hat m_2^2) 
\eeq
Using the identities,
\beq
(m_\pm^2 - \hat m_1^2) &=& -\tilde m^2 \pm m_R^2\nonumber\\
(m_\pm^2 - \hat m_2^2) &=& \tilde m^2 \pm m_R^2\,,
\eeq
the invertible unitary matrix of eigenvectors is easily calculated:
\beq
U = \frac{1}{\sqrt{m_p^4 +(\tilde m^2-m_R^2)^2}}\left( 
\begin{array}{cc}
m_p^2 &  (m_R^2 - \tilde m^2 )\\
- (m_R^2 - \tilde m^2 ) & m_p^2
\end{array}
\right)
:= 
\left(\begin{array}{cc}
\cos \theta & \sin \theta \\
-\sin \theta & \cos \theta \end{array} \right) \label{mat}
\eeq
Using (\ref{eom1}), (\ref{Otrans}) and (\ref{mat}) we obtain the
diagonalized form of the operator ${\cal O}$:
\beq
\label{op}
{\cal O}' = 
\left(
\begin{array}{cc}
-\boxx+\hat m_+^2 & 0\\
0      & -\boxx+\hat m_-^2
\end{array}
\right)\,.
\eeq
Note that when $\hat m_1 = \hat m_2$ (or $\tilde m =0$) this result reduces to a familiar $\pi/4$ rotation of field variables.  The matrix (\ref{mat}) becomes,
\beq
U~~\rightarrow~~
\frac{1}{\sqrt{2}} \left(
\begin{array}{cc}
1 & 1 \\
-1 & 1 
\end{array}
\right)
\eeq
which we use in (\ref{rot}) to obtain the simple rotational form of the transformation,
\beq
\Phi_1= \frac{1}{\sqrt 2}(\varphi_1 + \varphi_2)\nonumber\\
\Phi_2= \frac{1}{\sqrt 2}(\varphi_2 - \varphi_1)
\eeq

To rewrite the Green function in the eigenbasis of the coupled system
we solve the equation ${\cal O}' {\cal G}' = I$.  As before, we
separate the Green function into the sum of a particular solution and
a complementary function: ${\cal G}'=G' + 2\pi i {\cal F}'$ which
satisfy
\beq
{\cal O}' G' = I\,;~~{\cal O}' {\cal F'} =0 \label{e1}
\eeq
The inverse of the diagonal operator ${\cal O}'$ is trivial to
evaluate. We obtain
\beq
G' = \left( 
\begin{array}{cc}
G'_{11}& G_{12}'\\
G'_{21} & G_{22}'
\end{array}
\right) = \left( 
\begin{array}{cc}
\frac{1}{-\boxx+m_+^2}& 0\\
0 & \frac{1}{-\boxx+m_-^2}
\end{array}
\right).
\eeq
 In the coupled system, the diagonal form of the complementary function does not have to be preserved. However, we know that:
\beq
{\cal F}' = U {\cal F} U^{-1}. \label{x11}
\eeq
In addition, we know that ${\cal F}'$ must satisfy (\ref{e1}) which means that it must have the form,
\beq
{\cal F}' = \left(
\begin{array}{cc}
f'_{11}(k_0) \delta(\chi_+) & f'_{12} (k_0) \delta(\chi_+) \\
f'_{21}(k_0) \delta(\chi_-) & f'_{22} (k_0) \delta(\chi_-) 
\end{array}\right) \label{x2}
\eeq
We assume that the original $\hat{\cal F}$ is diagonal just prior to
the interaction begin turned on, so that if the interaction is turned
on suddenly, $\cal F$ remains diagonal:
\beq
{\cal F} = \left(
\begin{array}{cc}
f_{1}(k_0) \delta(\chi_1) & 0  \\
0  & f_{2} (k_0) \delta(\chi_2) 
\end{array}\right) \label{fdiag}
\eeq
where $\chi_1$ and $\chi_2$ are as yet undetermined mass shells.
Using (\ref{mat}) and (\ref{fdiag}) in (\ref{x11}) and comparing with
(\ref{x2}) we obtain the relationship between the components of $\cal
F$ and $\cal F'$:
\beq
\frac{f_{11}'}{\omega_+} &=& \frac{f_1}{ \omega_1}\,\cos^2\theta+\frac{ f_2}{\omega_2}\sin^2\theta \nonumber\\
\frac{f_{22}'}{\omega_-} &=& \frac{ f_{2}}{ \omega_2}\,\cos^2\theta+\frac{ f_{1}}{\omega_1}\sin^2\theta \nonumber\\
\frac{f_{12}'}{\omega_+} &=&  (\frac{f_{2}}{\omega_2}-\frac{ f_{1}}{\omega_1})\cos\theta\sin\theta \nonumber\\
\frac{f_{21}'}{\omega_-} &=&  (\frac{ f_{2}}{\omega_2}-\frac{ f_{1}}{\omega_1
})\cos\theta\sin\theta.
\label{distI}
\eeq

\section{Covariant analysis at $\overline t\not=0$}

Now we consider the behaviour of the system for $\overline t\not=0$.
We make use of the covariant method of \cite{burgess12} to solve for
the Wightman or complementary Green functions for the effective
theory.  The off-diagonal terms are irrelevant to further analysis;
they trivially satisfy the equations of motion by virtue of the fact
that their particular solutions are identically zero and their
complementary parts contain delta-function constraints.  For
notational simplicity we therefore refer to the diagonal components as
$G'_{11}:= G'_+$, $G'_{22}:= G'_-$, $f'_{11}:= f'_+$ and $f'_{22}:=
f'_-$.

Our strategy is to find the dispersion relation for the positive
frequency Wightman function. This dispersion relation does not have a
unique form: it depends on the ansatz used for the Green function
itself. By using a conformally covariant form for the ansatz and
introducing vector fields $A_\mu$, we obtain a form for the dispersion
relation which is independent of $\overline x$. This approach is
useful in two ways: it results in a neater form for the dispersion
relation, and it allows the internal conformal symmetry to be seen
directly and used to simplify calculations where the dispersion
relation is not $\overline x$ independent.  Note that we are not
claiming that the theory is conformally invariant. We are only noting
that an overall $\overline x$ scaling of the action can be used to
express any statistical theory in terms of dimensionless variables,
provided partial derivatives are replaced by covariant ones
\cite{burgess12}.  By formally introducing covariant connections, we
can exploit our ability to rescale a slowly varying non-equilibrium
theory so that it appears to be an equilibrium one, for the purposes
of calculation.

The positive frequency Wightman functions satisfy the
equations of motion:
\beq
\label{e2}
(-\stackrel{x}{\boxx}+m^2_a){G_a^{(+)}}'(x,x') &=& 0\,,
\eeq
where the index $a$ takes values $\{+,-\}$. 
We attempt to find solutions to these equations using the ansatz,
\beq
{G_a^{(+)}}'(x,x') &=& 2\pi i \int (d{\bf k})\;e^{ik\tilde x+ i\int^{\overline x}_0 A_{a\mu}\,dX^\mu} \frac{(1+f'_{a}(\omega_a(\overline x),\overline x))}{2\omega_a(\overline x)}\,.
\eeq
The formal dependence of the distribution functions $f'_{a}$ and on
the average coordinate $\overline x$ seems peculiar in a free theory,
where there is no apparent cause for a re-distribution of quanta
amongst different modes in either time or space. Such a change would
have to be represented by an additional source or sink (a potential),
which we do not have. However, there is nothing in the field equations
to exclude a combination of solutions which oscillates with a steady
periodic signature. Such a steady-state oscillation could be fed in as
a boundary condition. Another example would be the analysis of
low-order terms in an interacting theory, where the effective mass was
changing with time $m^2(t)$. Such a change makes no sense without a
broader theory to explain it, but as an effective model it can still
be analyzed in a consistent framework.

The purpose of the auxiliary field $A_\mu$ is as follows.  Since
external boundary conditions and renormalizations might impose a
time-dependence in the dispersion relation, one needs a generic term
$A_\mu$ to absorb the effect of the changing distribution. Its
introduction allows us to formulate the theory in terms of a
time-independent frequency, by moving the dependence of the average
coordinate into the phase.

Taking the derivative we obtain, 
\beq
\stackrel{x}{\partial_\mu} {G_a^{(+)}}'(x,x') &=& 2\pi i \int (dk) \Big[ 
i k_\mu + \frac{i}{2}A_{a\mu} 
+ i\int_0^{\overline x}(\tilde\partial_\mu A_{a\nu}) dX^\nu \nonumber\\
&-&\frac{i}{2}(\overline\partial_\mu \omega_a)\tilde t + \2 F'_{a\mu} - \2\Omega_{a\mu}
\Big] G_a^{(+)'}(k,\overline x)\,,
\eeq
where,
\beq
\Omega_{a\mu} = \partial_\mu \omega_a(\bar x)/\omega_a(\bar x)\,;~~~F'_{a\mu}= \frac{\partial_\mu f'_a}{1+f'_a} 
\eeq
and
\beq
\stackrel{x}{\partial}_\mu &=& \tilde \partial_\mu + \2 \overline \partial_\mu\nonumber\\
\stackrel{x}{\boxx} &=& \tilde\boxx + \tilde \partial_\mu\overline\partial^\mu + \frac{1}{4}\overline\boxx\,.
\eeq
Notice that the second term of the first line above behaves as an
auxiliary field. It is this term that will be absorbed by the
auxiliary field.

We proceed as follows.  We make the choice $A_{a\mu} = \tilde t \bar
\partial _\mu \omega_a$.  Substituting into $i\int_0^{\xbar} (\tilde
\partial_\mu A_{a\nu}) d \xbar '^\nu$ and integrating by parts we
obtain,
\beq
i\int_0^{\xbar} (\tilde \partial_\mu A_{a\nu}) d \xbar '^\nu=i\delta_{\mu}^{~0}[\omega_a(\xbar)-\omega_a(0)].
\eeq
When we substitute this result back into the dispersion relation we obtain,
\beq
\partial^x_\mu G_a^{(+)'} = 2\pi i \int (dk)\left[i(k_{a\mu} + \delta_{\mu}^{~ 0} [\omega_a(\xbar)-\omega_a(0)]) + \frac{1}{2}(F'_{a\mu} - \Omega_{a\mu})\right]G_a^{(+)'}
\eeq
We re-express this result covariantly by defining the vector
$k_{a\mu}(0) = \omega_a(0) \delta_{\mu 0} + k^i \delta_{i\mu}$ which
allows us to write
\beq
\partial_\mu G_a^{(+)'} = 2\pi i\int\,(dk)\,[ik_{a\mu}(0) + \frac{1}{2}(F'_{a\mu} - \Omega_{a\mu})]G_a^{(+)'}
\eeq
Performing another derivative and substituting into (\ref{e2}) we obtain,
\beq
-k_{a\mu}(0) k_a^\mu(0) - m_a^2 + \frac{1}{4}[\bar\partial^\mu(F'_{a\mu} - \Omega_{a\mu}) + (F'_a-\Omega_a)^2] +ik_a^\mu(0)(F'_{a\mu}-\Omega_{a\mu}) =0
\eeq
The first two terms give $-k_a(0)^2 - m_a^2= \omega_a^2(0) - {\bf k}^2
- m_a^2$ which we set to zero as an initial condition on
$\omega_a(\xbar)$. More precisely if $\omega_a$ is constant it must
equal the equilibrium value for the given mass.  Then we are left with
two equations
\beq
&& k_a^\mu(0)(F'_{a\mu}-\Omega_{a\mu})=0 \nonumber \\
&& \bar \partial^\mu(F'_{a\mu} - \Omega_{a\mu}) + (F'_a-\Omega_a)^2 =0 \label{second}
\eeq

\section{Solutions}

In order to look for solutions we set $\Omega_\mu=0$. The first part
of (\ref{second}) gives
\beq
k^\mu(0) \partial_\mu f'_a =0
\eeq
If we assume spatial homogeneity then this equation reduces to 
\beq
\partial_t f'_a=0\,.
\eeq
Now we want to ask what this result implies about the distribution
functions of the original degrees of freedom.  Using (\ref{distI}) we
obtain,
\beq
\partial_t f'_a=0~~\rightarrow~~
\left( \begin{array}{cc} \frac{\omega_+ \cos^2\theta}{\omega_1} & \frac{\omega_+ \sin^2\theta}{\omega_2} \\
\frac{\omega_- \cos^2\theta}{\omega_2}  & \frac{\omega_- \sin^2\theta}{ \omega_1} 
\end{array} \right)
\left( \begin{array}{c} \partial_t f_1 \\
\partial_t  f_2 \end{array}\right)
=0
\eeq
In general, this implies that the distribution functions in terms of
the original degrees of freedom must also be constant.  The only way
to obtain non-trivial solutions to this equation is to require that
the determinant of the matrix of coefficients must vanish.  This
requires
\beq
\cos^2\theta\,\sin^2\theta\,\omega_+^2 \omega_-^2 \left( \frac{1}{\omega_1^2} - \frac{1}{\omega_2^2} \right) =0\,.
\eeq
Since $\cos\theta$ and $\sin\theta$ can only vanish when the coupling,
$m_p$, vanishes (cf (\ref{neweigen})) the only possible solution to
this equation is for $\omega_1$ to equal $\omega_2$.

\section{A Generalizable Analysis}

An explicit rotation of variables is not the approach normally used in
solving field theories, since it does not readily generalize to
arbitrary higher order interactions.  An alternative method, which
leads to an effective theory for one of the systems can be found by
making a change of variable:
\beq
\Phi_2 &=& \phi_2 - \int (dx') G_2(x,x')\,J_2(x')\nonumber\\
       &=& \phi_2 - m_p^2 \int (dx') G_2(x,x')\,\Phi_1(x').\label{shift}
\eeq
The shift involves a non-local Green function, which is assumed to satisfy
\beq
(-\boxx + \hat m_2^2) G_2(x,x') = \delta (x,x').\label{m2}
\eeq
The boundary conditions on $G_2(x,x')$ are not determined by this
equation or by the Schwinger action principle itself.  To determine
these boundary conditions we use the fact that the interacting fields
$\Phi_1$ and $\Phi_2$ are influenced by one another only after they
are coupled at time $\overline t=0$. This physical consideration leads
us to choose the retarded Green function.

Making the shift of variables (\ref{shift}) in  (\ref{action}) gives,
\beq
S_{\rm tot} &=& \int (dx) \Big\{
\Phi_1^*(x) (-\boxx + \hat m_1^2) \Phi_1(x) - m_p^4 \int (dx') \Phi_1^*(x)G_2(x,x')\Phi_1(x') \nonumber\\
&+& \phi_2^*(x)(-\boxx+\hat m_2^2)\phi_2(x)
\Big\}\label{effact}
\eeq
We are left with a non-local interaction in terms of the original
variables.  The resulting equations of motion have the matrix form,
\beq \tilde{\cal O} \left( 
\begin{array}{c}
\Phi_1 \\ \phi_1
\end{array}
\right) =0 
\eeq
with
\beq
\tilde{\cal O} = 
\left(
\begin{array}{cc} 
-\boxx + \hat m_1^2 - \frac{m_p^4}{-\boxx + m_2^2} & 0 \\
0 & -\boxx + \hat m_2^2 
\end{array}
\right)
\label{eom2}
\eeq
To find the particular part of the corresponding Green function we
need to invert the matrix $\tilde{\cal O}$.  We obtain,
\beq
\tilde{\cal O}^{-1} = \frac{1}{{\rm det} \tilde{\cal O}} \left(
\begin{array}{cc}
-\boxx + \hat m_2^2 & 0 \\
0 & -\boxx + \hat m_1^2 - \frac{m_p^4}{-\boxx + m_2^2} 
\end{array}
\right)\,.
\eeq
Using
\beq
{\rm det}\tilde{\cal O} = (-\boxx + \hat m_1^2)(-\boxx + \hat m_2^2) - m_p^4 = (-\boxx + m_+^2)(-\boxx + m_-^2)
\eeq
it is clear that the Green function has poles at the excitation masses
$m_+$ and $m_-$. This result agrees with the result found in the
previous section: the true mass shell of the excitations is given by
the eigenmasses of the system.

In order to find the complementary part of the Green function we
follow the method of the previous section.  We define the matrix
\beq
\Lambda = \left( 
\begin{array}{cc}
1 & 0\\
\frac{-m_p^2}{-\boxx+\hat m_2^2} & 1
\end{array}
\right)\label{lam}
\eeq
The equation that is analogous to (\ref{rot}) is then
\beq
\Phi = \Lambda \phi \,.
\eeq
We can obtain the diagonalized matrix $\tilde{\cal O}$ given in
(\ref{eom2}) by performing the transformation on ${\cal O}$ analogous
to (\ref{Otrans}):
\beq
\tilde{\cal O} = (\Lambda^\dagger)^{-1} {\cal O} \Lambda^{-1} \,.
\eeq
Using ${\cal O}$ as defined in (\ref{eom1}) we reproduce
(\ref{eom2}). The complementary piece of the Green function is
obtained from the equation analogous to (\ref{Otrans}):
\beq
\tilde{\cal F} = (\Lambda^\dagger)^{-1} {\cal F} \Lambda^{-1}\,. \label{x111}
\eeq

To proceed we rewrite (\ref{eom2}) in the form  
\beq
\tilde{\cal O} = \left(
\begin{array}{cc}
\frac{1}{-\boxx + \hat m_2^2}(-\boxx + m_+^2)(-\boxx + m_-^2) & 0 \\
0 & (-\boxx + m_2^2) 
\end{array}
\right)\,,
\eeq
and impose $\tilde{\cal O} \tilde{\cal F}=0$ by writing
\beq
\tilde{\cal F} = \left(
\begin{array}{cc}
\tilde f_{11} \delta(\chi_\pm) & \tilde f_{12} \delta(\chi_\pm) \\
\tilde f_{21} \delta(\hat\chi_2) & \tilde f_{22} \delta(\hat\chi_2) 
\end{array}
\right)
\label{x14}
\eeq
where $\delta(\chi_\pm)$ indicates any linear combination of
$\delta(\chi_+)$ and $\delta(\chi_-)$.  Using (\ref{fdiag}) and
(\ref{lam}) in (\ref{x111}) and comparing with (\ref{x14}) we obtain,
\begin{eqnarray}
\tilde f_{22}\delta(\hat\chi_2) &&=  f_2 \delta(\chi_2)\nonumber \\
\tilde f_{11} \delta(\chi_\pm) &&=  f_1 \delta(\chi_1) + \frac{m_p^4}{(-\boxx + \hat m_2^2)^2} f_2 \delta(\chi_2) \label{r1}
\end{eqnarray}
In order to study the time dependence of the distribution functions
$\tilde{\cal F}$ we proceed as in the previous section. The
$\phi_2(x)$ Wightman function obeys the equation of motion,
\beq
(-\stackrel{x}{\boxx} + \hat m_2^2){\tilde G_2^{(+)}}(x,x')=0, \label{s1}
\eeq
and its dispersion constraint is the free mass-shell
\beq
\hat\chi_2 = k^2 + \hat m_2^2 = 0. 
\eeq
We postulate an ansatz of the form,
\beq
{\tilde G_2^{(+)}}(x,x') = 2\pi i \int\; (d{\bf k}) 
e^{ik\tilde x+i\int_0^{\overline x} A_{2\mu}\,dX^\mu} 
\frac{(1+\tilde f_{22}(\omega_{22}(\overline x),\overline x))}{2\omega_{22}(\overline x)}.
\eeq
 Substituting into (\ref{s1}) gives,
\beq
\partial_t \tilde f_{22}=0
\eeq
Using (\ref{r1}) we get $\partial_t  f_2 =0$ or $ f_2 =$ constant.  

The $\Phi_1(x)$ Wightman function is more complex; it satisfies the
equation of motion,
\beq
\label{eom3}
(-\stackrel{x}{\boxx} + \hat m_1^2){\tilde G_1^{(+)}}(x,x') - m_p^4 \int (dx'') G_{2}(x,x''){\tilde G_1^{(+)}}(x'',x')=0\,.
\eeq
To simplify this expression we operate on the left with
$(-\stackrel{x}{\boxx} + \hat m_2^2)$ which gives,
\beq
[(-\stackrel{x}{\boxx} + \hat m_2^2)(-\stackrel{x}{\boxx} + \hat m_1^2) - m_p^2 ] \tilde G_1^{(+)} = (-\stackrel{x}{\boxx} + m_+^2)(-\stackrel{x}{\boxx} + m_-^2) \tilde G_1^{(+)} =0 \label{x15}
\eeq
We postulate an expression for ${\tilde G_1^{(+)}}(x,x')$ of the form,
\beq
{\tilde G_1^{(+)}}(x,x') = 2\pi i \int\; (d{\bf k}) 
e^{ik\tilde x+i\int_0^{\overline x} A_{1\mu}\,dX^\mu} 
\frac{(1+\tilde f_{11}(\omega_{11}(\overline x),\overline x))}{2\omega_{11}(\overline x)}.
\label{x16}
\eeq
Substituting (\ref{x16}) into (\ref{x15}) we could solve for the
dispersion relation, as in the previous section.  The formalism
discussed in this section could be used to study a system with a more
complicated non-linear interaction, which would lead to non-trivial
solutions.

\section{Conclusion}

We have studied a system which consists of two initially independent
scalar systems that are minimally coupled at some initial time.  We
have used the machinery of Green functions and effective field theory
to study the time evolution of the statistical distribution
functions. This problem is an inherently non-equilibrium one.  We have
discussed how to parameterize such an equilibrating system, and how to
perform calculations when equations of motion and thus dispersion
relations change with time.  Building on the earlier work in
\cite{calzetta1,burgess12} we have adopted a covariant approach to
handle time dependent dispersion relations and used the underlying
conformal structure to show the consistency of working with
perturbation theory in which dispersion relations are
time-independent. We have shown that by formally introducing covariant
connections, one can rescale a slowly varying non-equilibrium theory
so that it appears to be an equilibrium one, for the purposes of
calculation. This approach turns out to be of only formal utility in
the example provided, but can be generalized to more complex and
realistic models.  The simple minimally coupled system studied here
does not reach a conventional kind of macroscopic equilibrium.  In
order to study more conventional equilibration, in which the
statistical properties are related to the mechanical interactions, we
would need to include higher order interactions.


\vspace*{1cm}

\Large

\noindent {\bf Acknowledgment}

\normalsize

\noindent This work is supported by NATO collaborative research grant CRG950018.

\bibliographystyle{unsrt}

\end{document}